\documentclass[preprint]{aastex61}
\usepackage{amsmath,amstext}
\usepackage{graphicx}

\received{****}
\revised{****}
\accepted{****}

\begin{document}

\title{The Parametric Decay Instability of Alfv\'en waves in Turbulent Plasmas and the Applications in the Solar Wind}

\correspondingauthor{Chijie Xiao}
\email{cjxiao@pku.edu.cn}

\author{Mijie Shi}
\affiliation{State Key Laboratory of Nuclear Physics and Technology, Fusion Simulation Center, School of Physics, Peking University, Beijing 100871, China}
\affiliation{Los Alamos National Laboratory, Los Alamos, New Mexico 87545, USA}

\author{Hui Li}
\affiliation{Los Alamos National Laboratory, Los Alamos, New Mexico 87545, USA}

\author{Chijie Xiao}
\affiliation{State Key Laboratory of Nuclear Physics and Technology, Fusion Simulation Center, School of Physics, Peking University, Beijing 100871, China}

\author{Xiaogang Wang}
\affiliation{State Key Laboratory of Nuclear Physics and Technology, Fusion Simulation Center, School of Physics, Peking University, Beijing 100871, China}
\affiliation{Department of Physics, Harbin Institute of Technology, Harbin 150001, China}



\begin{abstract}

We perform three dimensional (3D) ideal magnetohydrodynamic (MHD) simulations to study the parametric 
decay instability of Alfv\'en waves in turbulent plasmas and explore its possible applications in the solar wind. 
We find that, over a broad range of parameters in background turbulence amplitudes, the parametric decay instability of an Alfv\'en wave with various amplitudes can still occur, though its growth rate in turbulent
plasmas tends to be lower than both the theoretical linear theory prediction and that in the non-turbulent situations. 
Spatial - temporal FFT analyses of density fluctuations produced by the parametric decay instability 
match well with the dispersion relation of the slow MHD waves. This result may provide an
explanation of the generation mechanism of slow waves in the solar wind observed at 1 AU. It further 
highlights the need to explore the effects of density variations in modifying the turbulence properties
as well as in heating the solar wind plasmas.
\end{abstract}

\keywords{Alfv\'en waves --- 
slow waves --- parametric instability --- solar wind}



\section{Introduction} \label{sec:intro}

The solar wind plasma is a natural laboratory to study processes relevant to space physics, plasma physics, 
as well as astrophysics \cite[e.g.,][]{1995SSRv...73....1T,2013LRSP...10....2B,2016SSRv..201...55A}. 
Spacecraft measurements show that the evolution of the solar wind is not adiabatically expanding, 
instead, additional heating effects are required during the expansion of the solar wind 
\cite[e.g.,][]{1995GeoRL..22..325R,2003GeoRL..30.1206R}. Numerous processes can heat the 
solar wind plasma, for example: dissipation of waves, stream interactions and shocks, pickup ions, etc. 
MHD waves, the low frequency (lower than ion cyclotron frequency $\Omega_{ci}$) plasma waves, 
could play a significant role in heating both solar wind and solar corona 
\citep{2009ApJ...707.1659C,2010LRSP....7....4O,2012ApJ...754...92C}. Among the three MHD waves, 
Alfv\'en waves have received the greatest attention because it is observed as the dominant component 
in the solar wind \citep{1971JGR....76.3534B,1995SSRv...73....1T,2016ApJ...816L..24Y}. There are 
some evidences that Alfv\'en waves also exist in the solar chromosphere and solar corona 
\citep{2007Sci...318.1574D,2009A&A...497..525H,2011Natur.475..477M}. Alfv\'enic turbulence cascade 
has been well studied, where counter propagating Alfv\'en waves cascade and transfer energy from 
large scales to small scales and dissipate into heat 
\citep{1984JGR....89.9695T,1995ApJ...438..763G,2009ApJS..182..310S}. 

The other two MHD waves, slow wave and fast wave, seem unlikely to exist in the solar wind 
plasma owing to the strong Landau damping 
\citep{1966PhFl....9.1483B,1972JPlPh...8..197B,1979JGR....84.4459B}. However, recent observations 
have shown the existence of slow waves. By analyzing the correlation between proton density and 
field aligned magnetic field, as well as the correlation's dependence on ion plasma $\beta$, 
\cite{2012ApJ...753L..19H} showed the compressible components of solar wind turbulence are 
kinetic slow waves. \cite{2013ApJ...774...59Y} found that some of the pressure balanced structures 
\citep{1970SoPh...15...61B} in the solar wind are driven by oblique propagating slow waves, 
because oblique propagating slow waves suffer less from Landau damping \citep{1979JGR....84.4459B} 
and may survive in the solar wind. \cite{2015ApJ...815..122S} identified some slow wave cases in 
the solar wind by dispersion relation comparison.  \cite{2015ApJ...813L..30H} observed the interaction 
between Alfv\'en waves and slow waves, and their effects on ion heating. Slow waves are also 
observed in the bursty flow of plasma sheet \citep{2016GeoRL..43.1854W}. Recently, \cite{2017arXiv170303040V} found that the large scale compressible fluctuations in the solar wind are more like MHD slow modes than kinetic slow modes. As far as we know, 
the origin of slow waves in the solar wind is not clear. In this paper, we will propose a generation 
mechanism of slow waves in the solar wind turbulence.

In compressible plasmas, Alfv\'en waves will be subject to the parametric decay instability (PDI), 
during which forward propagating compressive fluctuations are produced and backward propagating 
Alfv\'en waves are also generated. Studies of PDI have a long history. 
For example, with a parallel propagating circularly polarized Alfv\'en wave 
in one-dimension, \cite{1978ApJ...224.1013D} and \cite{1978ApJ...219..700G}  derived a general 
dispersion relation for the growing compressive mode. \cite{1994JGR....9923431H} provided a 
derivation of different kinds of parametric instabilities. Dispersive effects were also introduced to 
study this process \citep{1986JGR....91.5617W}. Many simulations for both monochromatic and 
non-monochromatic Alfv\'en waves with different polarizations have shown that the PDI of Alfv\'en waves 
is a robust process \cite[e.g.,][]{2000PhPl....7.2866M,2001A&A...367..705D,2001GeoRL..28.2585Z,2010GeoRL..3720101M,2013PhPl...20g2902G,2015JPlPh..81a3202D}. 
Furthermore, the PDI of Alfv\'en waves has been considered as an important mechanism to 
generate counter-propagating Alfv\'en waves, which are necessary for Alfv\'enic turbulence cascade, 
as well as to explain the decrease of cross helicity with the increasing heliosphere distance 
\citep{1987JGR....9212023R,2000JGR...10512697B}. Comparatively, however, the properties and 
consequences of the density fluctuations generated during the PDI of Alfv\'en wave are not studied in detail. 
It is possible that the density fluctuations may be partially responsible for producing the compressible 
slow modes in the solar wind.

In this work, we examine the applicability of the PDI of Alfv\'en waves in the solar wind by studying
the propagation and dynamics of a single Alfven wave in the turbulent background. 
By comparing the PDI of Alfv\'en waves in turbulent and non-turbulent situations, we find that the 
instability can still occur. In Section \ref{sec:sims}, we provide the simulation setup and results. In Section 
\ref{sec:sum} 
we give the summary and discuss the possible applications.

\section{Simulation Results}
\label{sec:sims}
\subsection{Simulation Setup}
\label{sec2.1}

Simulations are performed using the ideal MHD module of PLUTO code \citep{2007ApJS..170..228M}. 
Two sets of simulations are performed which are categorized as  
turbulence simulations and non-turbulence simulations. 
The simulation domain is a 3D box elongated along the background magnetic field: 
$\boldsymbol{B}_0=B_0 \hat{\boldsymbol{x}}$. Simulation domain is $\left[8\pi,2\pi,2\pi\right]$ and 
the grid size is $\left[1024,256,256\right]$ in $x,y,z$ direction, respectively. Periodic boundary 
conditions are applied in all directions. In normalized units, $B_0=1,\rho_0=1$, Alfv\'en speed 
$v_A=1$, sound speed $c_s=\sqrt{\gamma\beta/2}$, where $\beta$ is the ratio between thermal 
pressure and magnetic pressure, and $\gamma=5/3$ is the adiabatic index. 
Time is normalized by $\tau_A$, where $\tau_A = 8\pi/v_A$ is Alfv\'en crossing time of the simulation 
domain along the $\hat{\boldsymbol{x}}$ direction. 

\begin{figure}
	\begin{centering}
		\includegraphics[width=\linewidth]{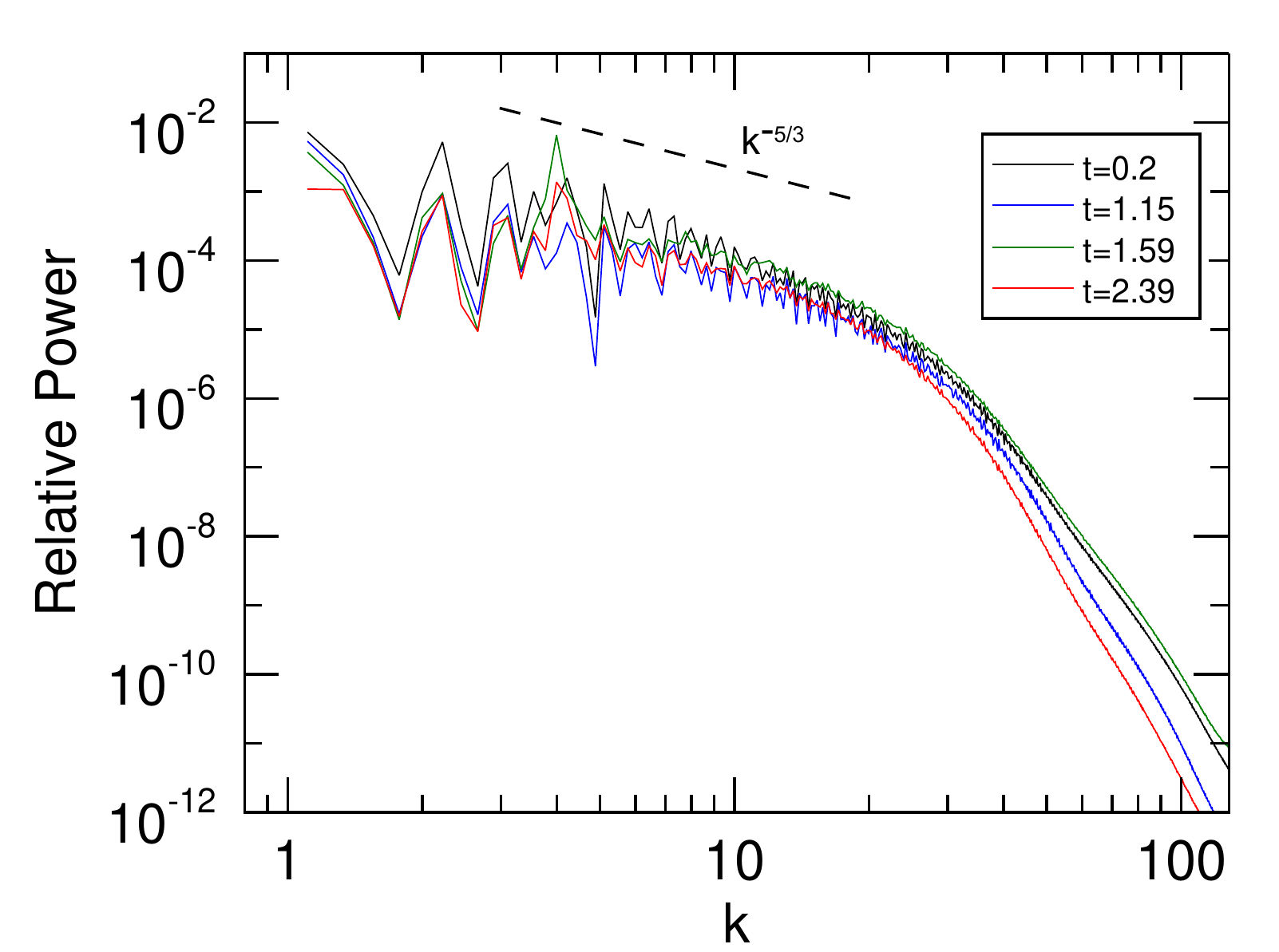}
		\caption{Power spectrum of $B_z$ component at different times of one turbulence simulation ($\delta B_{turb}=\delta v_{turb} = 0.15$). The spectrum is normalized by the grids size. At $t=1.2$, an Alfv\'en wave with $\delta B =0.2$ is injected into the simulation domian.}
		\label{fg1}
	\end{centering}
\end{figure}

For the turbulence simulations, we adopt a two-step process. First, we produce a turbulent background
in the simulation domain following the procedure described in \cite{2015PhPl...22d2902M}.  
Six linearly polarized Alfv\'en waves are injected into the domain, with their magnetic field 
and velocity perturbations in the following form, 
\begin{eqnarray}
\delta\boldsymbol{B}_{turb} & = &\sum\limits_{j,k}\delta B_{turb}\cos(jk_xx+lk_zz+\phi_{j,l})\hat{\boldsymbol{y}}
+\sum\limits_{m,n}\delta B_{turb}\cos(mk_xx+nk_yy+\phi_{m,n})\hat{\boldsymbol{z}}
\label{eq1}\\
\delta\boldsymbol{v}_{turb} & = & -\sum\limits_{j,k}sgn(j)\delta v_{turb}\cos(jk_xx+lk_zz+\phi_{j,l})\hat{\boldsymbol{y}}
-\sum\limits_{m,n}sgn(m)\delta v_{turb}\cos(mk_xx+nk_yy+\phi_{m,n})\hat{\boldsymbol{z}}\label{eq2}
\end{eqnarray}
where $(j,l)=(1,1),(1,2), (-2,3)$, $(m,n)=(-1,1),(-1,-2),(2,-3)$, and $k_{x,y,z}=2\pi/L_{x,y,z}$. 
Random phases are given by $\phi$, and $sgn$ is the sign function. There are three waves 
propagating obliquely in the positive $\hat{\boldsymbol{x}}$ and negative $\hat{\boldsymbol{x}}$  direction, 
respectively. The amplitudes $\delta B_{turb}$ and $\delta v_{turb}$ characterize the initial strength 
of the fluctuations and they also determine the magnitude of turbulence at later times when an Alfv\'en wave 
is injected. 
These counter-propagating Alfv\'en waves will cascade and form a decaying turbulent state. 
The turbulence is typically established after one Alfv\'en time 
\cite[see][for details about the turbulence properties]{2015PhPl...22d2902M}. 
Second, 
at $t = 1.2$, a circularly polarized Alfv\'en wave is injected into the domain with
\begin{eqnarray}
\delta\boldsymbol{B}_{cir} &=& \delta Bcos(4x)\hat{\boldsymbol{y}}+\delta Bsin(4x)\hat{\boldsymbol{z}}
\label{eq3}\\
\delta\boldsymbol{v}_{cir}&=& -\delta\boldsymbol{B}_{cir} \label{eq4}
\end{eqnarray}
where $\delta B$ is the amplitude of the injected Alfv\'en wave and it has a wavenumber $k_{x0} = 4$. 
Simulations are typically extended to 4 Alfv\'en time to study the evolution of PDI.  The power spectrum of $B_z$ at different times are shown in Figure \ref{fg1}. Before the Alfv\'en wave injection, the system has decayed to a turbulent state (blue line). After the injection of an Alfv\'en wave, a discrete mode at $k=4$ with a relatively high power (corresponding to the injected Alfv\'en wave) is now visible in the spectrum (green line).

To our knowledge, all previous PDI studies were in the non-turbulence situation 
\cite[e.g.][]{2001A&A...367..705D}. Typically, a circularly polarized Alfv\'en wave described by 
Equations \ref{eq3} and \ref{eq4} is injected into the simulation domain with a
uniform background in density, pressure
and magnetic field, plus a very small random density perturbations 
($\delta \rho/\rho_0 \sim 10^{-2}$). This is what we used as well in non-turbulence simulations.

\subsection{Parametric Decay Instability in Turbulence}

The PDI of Alfv\'en waves is characterized by an exponential growth stage of the
density fluctuations before saturation. In the one dimension (1D) limit, 
theoretical growth rates can be obtained from solving the dispersion relation derived by  
\cite{1978ApJ...224.1013D} and \cite{1978ApJ...219..700G}:
\begin{equation}
(\omega^2-\beta k^2)(\omega+k+2)(\omega+k-2)(\omega - k)\\
=\eta^2k^2(\omega^3+k\omega^2-3\omega+k)
\label{eq5}
\end{equation}
where $\omega$ and $k$ are normalized by the frequency and wavenumber of the injected 
Alfv\'en wave and $\eta=\delta B/B_0$. Note that $\beta$ in Equation \ref{eq5} is defined as 
$c_s^2/v_A^2$. Furthermore, for a given $\beta$ and $\eta$, in multi-dimensional studies 
such as ours, we find that many oblique modes with different $\omega$ and ${\bf k}$ are excited due to PDI. 

\begin{figure}
	\begin{centering}
		\includegraphics[width=\linewidth]{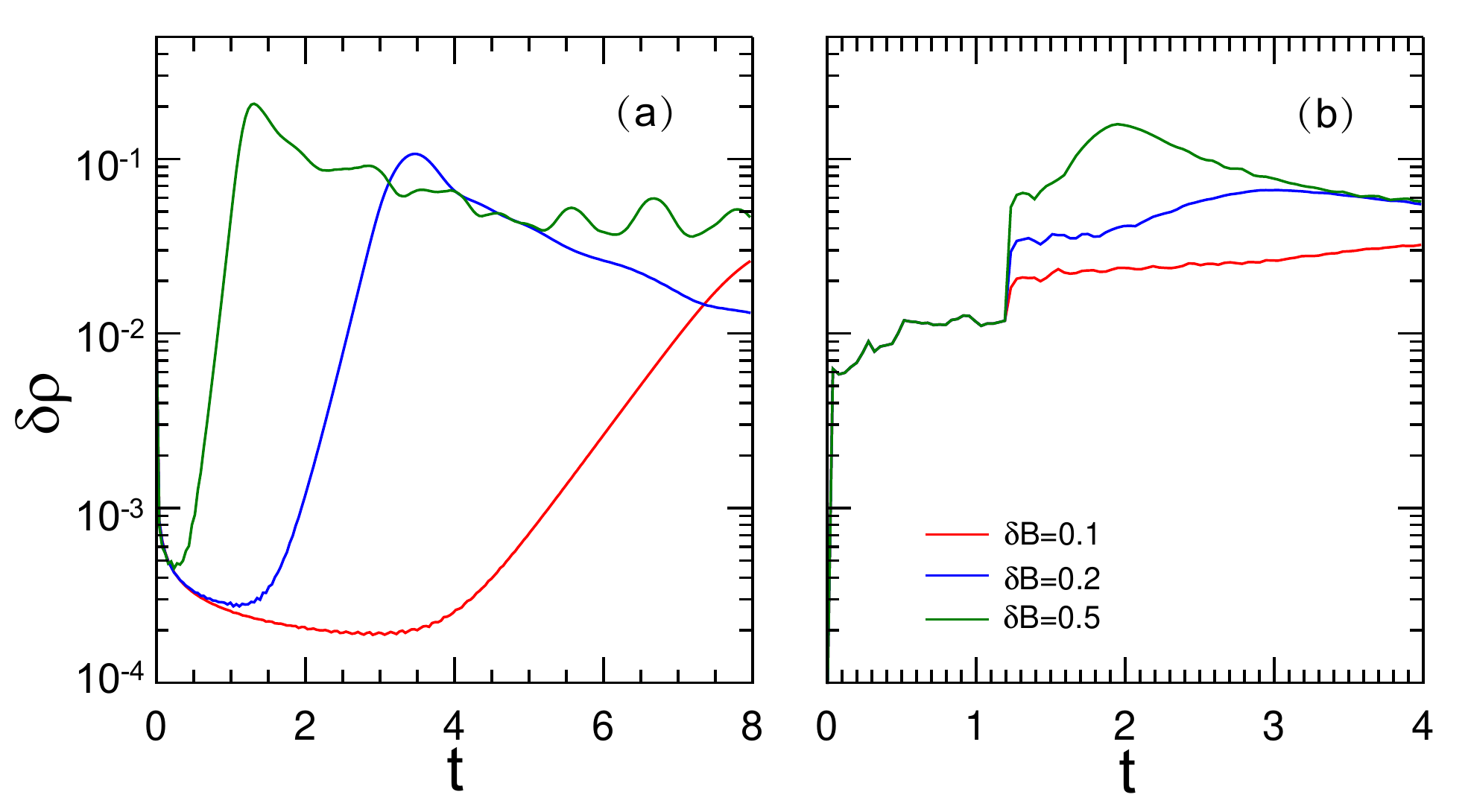}
		\caption{Evolution of root mean square (RMS) density fluctuation for (a) non-turbulence simulations and (b) turbulence simulations ($\delta B_{turb}=\delta v_{turb} = 0.05$). Red, blue, and green lines are for $\delta B=0.1, 0.2,$ and $0.5$, respectively. Plasma $\beta = 0.5$ for all simulations. At $t=1.2\tau_A$, an Alfv\'en wave is injected into the domian in turbulence simulations.}
		\label{fg2}
	\end{centering}
\end{figure}

Figure \ref{fg2} (a) shows, in the non-turbulence simulations, 
the root mean square (RMS) density fluctuation evolution for different amplitudes $\delta B$
of the injected Alfv\'en waves. The plasma $\beta = 0.5$ for all the simulations. 
An exponential growth stage is clearly seen for all simulations in Figure \ref{fg2} (a), 
with the growth rate becoming larger when $\delta B$ is larger. Note that even though the initial 
random density fluctuation amplitude is $\sim 0.01$, it quickly drops below $10^{-3}$, out of which
PDI gradually grows. The growth of the RMS density is due to a single growing mode.

For the turbulence simulations 
with $\delta B_{turb}=\delta v_{turb} = 0.05$, the RMS density fluctuation evolution is shown in 
Figure \ref{fg2} (b). The density fluctuation increases rapidly and remains nearly constant at $\sim 0.01$
during the cascade of Alfv\'en waves to develop turbulence. At $t = 1.2$, when a circularly 
polarized Alfv\'en wave is injected into the domain, the density fluctuation shows a sudden jump (most likely
due to the ponderomotive effect of the injected wave), followed by an exponential growth stage. 
This is the signature that the injected Alfv\'en wave undergoes the parametric decay instability. 
The dependence of the PDI's growth rate on the amplitude of the injected waves can be seen 
from the slope changes during the exponential stage.

\begin{figure}
	\begin{centering}
		\includegraphics[width=\linewidth]{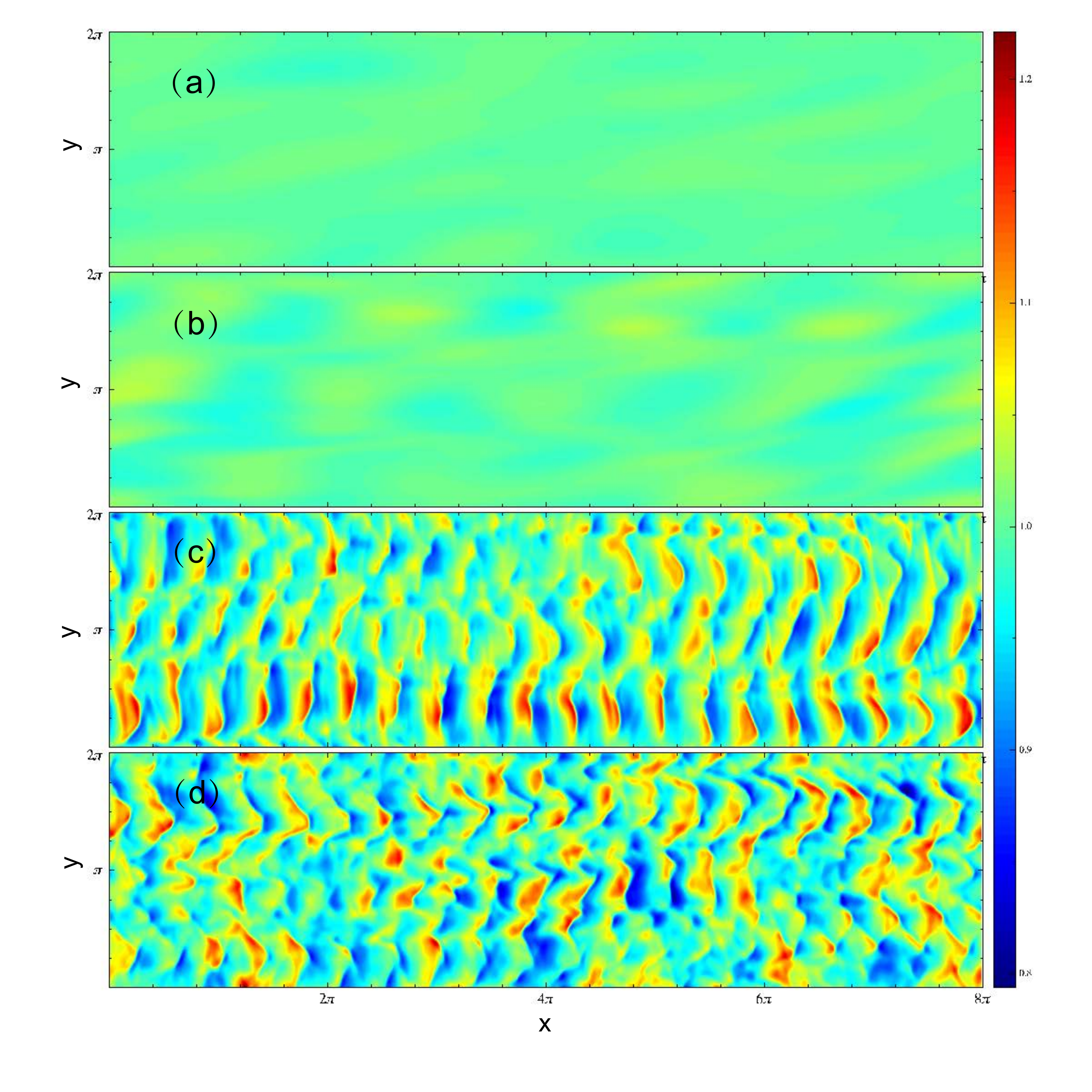}
		\caption{Evolution of density distribution in the $\{x,y\}$ plane ($z=\pi$) for a turbulence case
			with $\delta B = 0.2$ at (a) $t=0.08$, (b) $t=1.15$, (c) $t=2.39$, and (d) $t=3.98$. An Alfv\'en wave is injected at $t=1.2$.}
		\label{fg3}
	\end{centering}
\end{figure}

Figure \ref{fg3} shows the evolution of density distribution in the $\{x,y\}$ plane ($z=\pi$) for a turbulence case
with $\delta B = 0.2$ in Figure \ref{fg2}. After the injection of the circularly polarized Alfv\'en wave ((c) and (d) in Figure \ref{fg3}), the density mode shows a dominant mode with $k_x = 5$ and
$k_y = k_z = 0$, which is exactly what the 1-D linear theory
of PDI predicts with this parameter set . In addition, oblique slow modes with non-zero $k_y$ and $k_z$ are
also developed, though with smaller growth rates. 
These results demonstrate that the PDI of a single Alfv\'en wave still occurs with the background turbulence, 
with both parallel and obliquely propagating slow modes (see below for detailed mode analysis). 

\begin{figure}
	\begin{centering}
		\includegraphics[width=\linewidth]{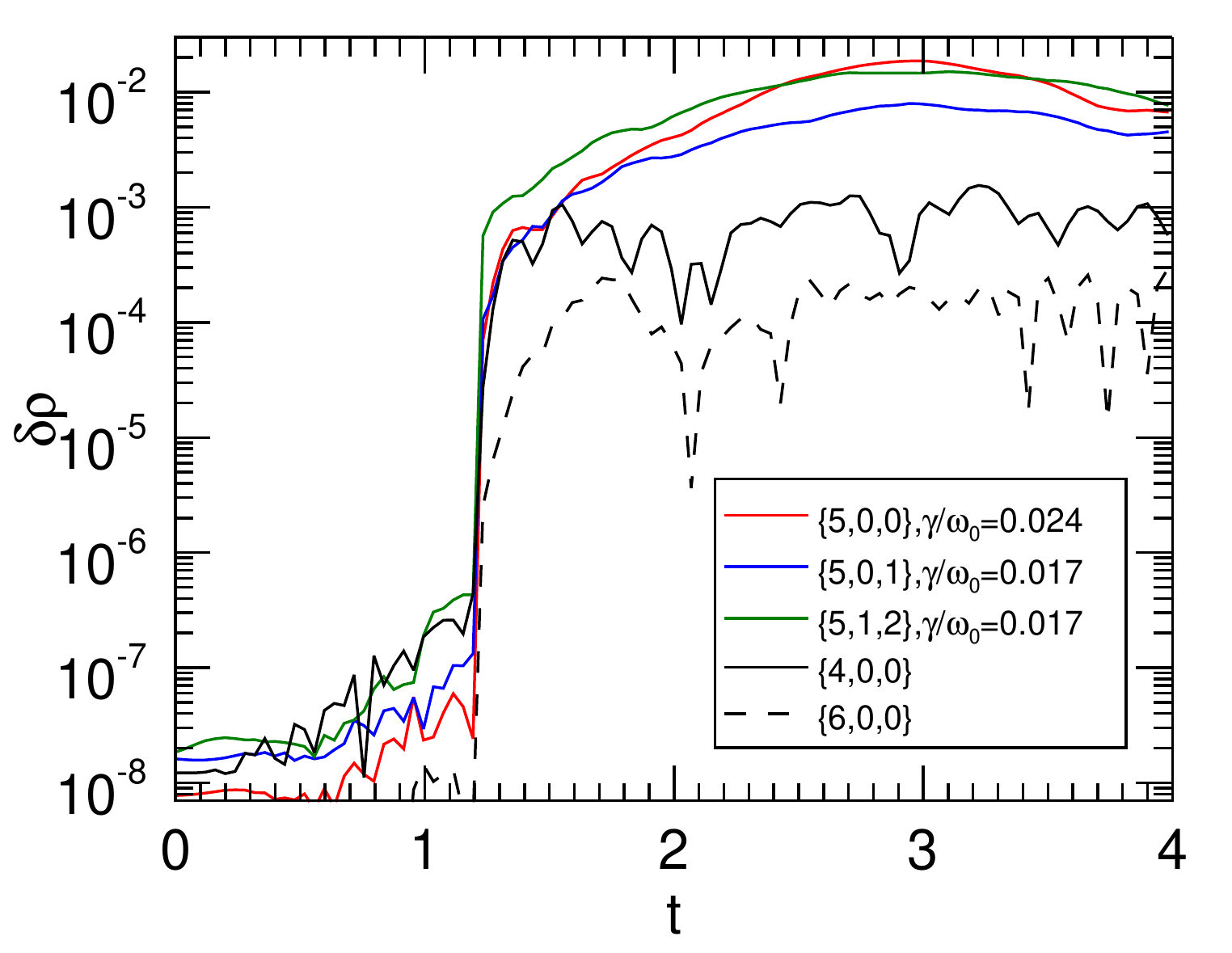}
		\caption{Evolution of RMS density fluctuation of three PDI modes $\{k_x,k_y,k_z\}=\{5,0,0\},\{5,0,1\}$, and $\{5,1,2\}$ as well as two other modes $\{k_x,k_y,k_z\}=\{4,0,0\},\{6,0,0\}$. The growth rates of PDI modes are labeled in the legends. A circularly polarized Alfv\'en wave with $\delta B =0.2$ is injected at $t=1.2$.}
		\label{fg4}
	\end{centering}
\end{figure}

To examine the mode growth in detail, we have carried out the FFT analysis of the density fluctuations
and found that some modes exhibit exponential growth. Three of them ($\{k_x, k_y, k_z\} = \{5,0,0\},\{5,0,1\},\{5,1,2\}$) are shown in Figure \ref{fg4}. We calculate the growth rates of the three modes in the time range between $t=1.6$ and $t=2.4$. The growth rates for the three modes are $0.024$, $0.017$, and $0.017$, respectively, as shown in the legend of Figure \ref{fg4}. In particular, the mode $\{k_x, k_y, k_z\} = \{5, 0,0\}$ (predicted by the linear theory) has the highest growth rate
and is responsible for the most of the density RMS growth seen in Figure \ref{fg2} (b). Figure \ref{fg4} also shows two other modes ($\{k_x, k_y, k_z\} = \{4,0,0\},\{6,0,0\}$), which have less energy than these PDI modes.


To quantitatively study the influence of turbulence on the growth rate of PDI, we have investigated 
two types of situations: one has the same background turbulence but different injected single Alfv\'en wave 
amplitudes and the other is the same injected Alfv\'en wave 
amplitudes but different background turbulence levels. We now discuss them in detail. 

First, we present results from a suite
of 3D simulations with the same turbulence background and the same single Alfv\'en wave injection time ($t=1.2$)
except that we vary the injected Alfv\'en wave amplitude, ranging from $\delta B/B = 0.1$ to $0.5$. 
The plasma $\beta = 0.5$. 
The 1D linear theory predicts that the PDI mode with an integer wave-number should have $k_x = 5$ for 
all these parameters. From our non-linear 3D turbulence simulations, 
we confirm that indeed in all cases the maximum growth mode has $k_x =5$. 
We then utilize the FFT analysis to select all the density modes with $k_x=5$ and integrate over 
$k_y$ and $k_z$. The growth rates are obtained during the exponential growth stage of their time histories.

\begin{figure}
	\begin{centering}
		\includegraphics[width=\linewidth]{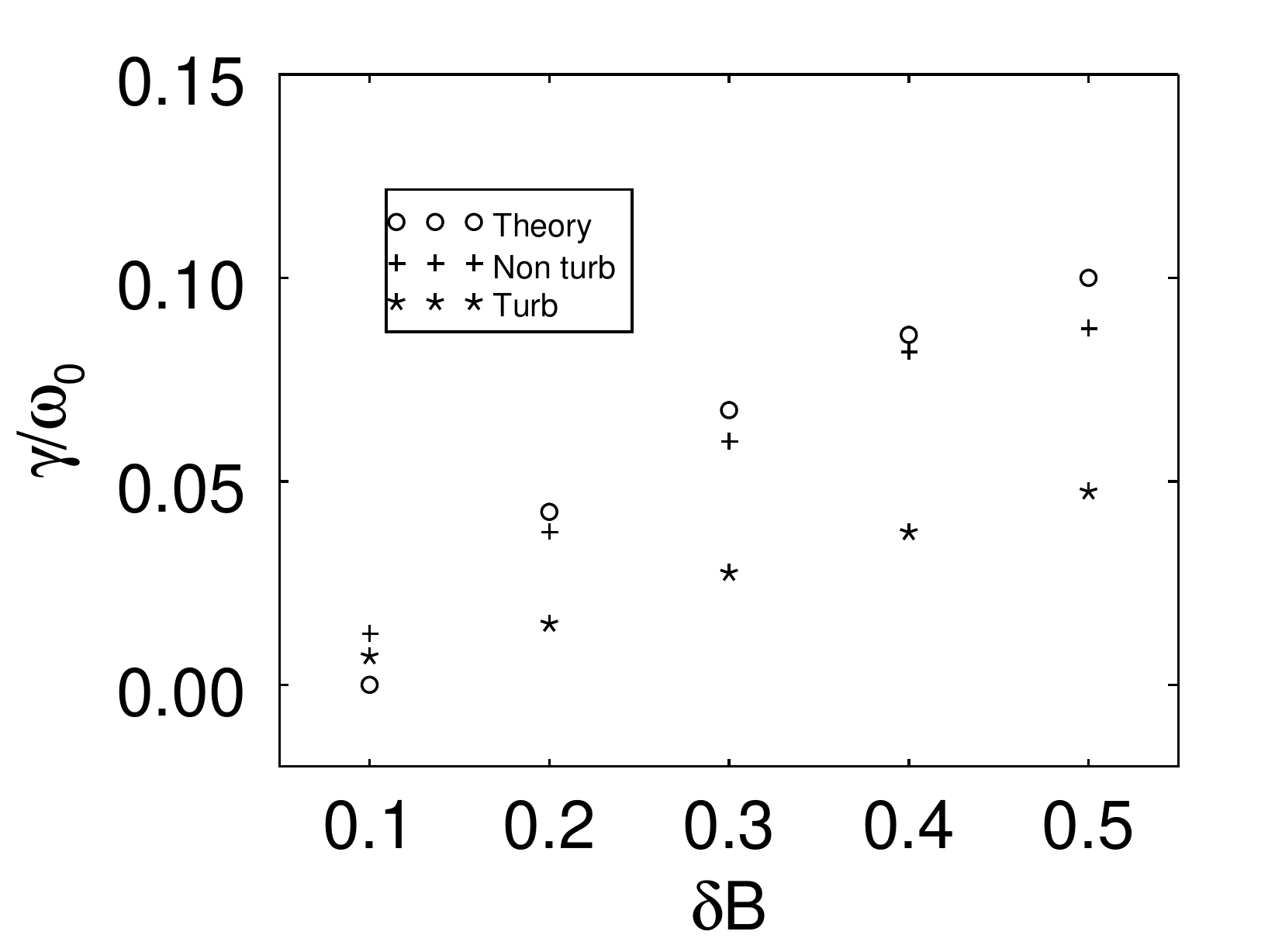}
		\caption{Growth rate comparison of PDI modes for turbulence simualtions ($\star$) ($\delta B_{turb}=\delta v_{turb} = 0.05$), non-turbulence simualtions (+), and theoretical predictions (o) from Equation \ref{eq5}, with different amplitudes of circularly polarized Alfv\'en waves. The plasma $\beta = 0.5$.}
		\label{fg5}
	\end{centering}
\end{figure}

Figure \ref{fg5} shows the growth rate of PDI for different $\delta B$ from three different situations: the 1-D linear
theory, non-turbulence simulations and turbulence simulations with $\delta B_{turb}=0.05$. 
The growth rates of non-turbulence simulations match the theoretical results with an error of less than 10\%. 
The growth rates of turbulence simulations, however, are about half of the theoretical results.
This suggests that the background turbulence has decreased the growth rate of PDI of Alfv\'en waves. 

One interesting point worth noting is that, at $\delta B = 0.1$, the linear theory predicts there is no 
growth rate at the integer mode $k = 5$, though finite PDI growth is predicted for (non-integer) 
wavenumber very close to $k=5$.  Interestingly, both the turbulence and non-turbulence simulations 
show the growth of the $k_x=5$ mode. We speculate that this is because the turbulence and/or the background
fluctuations give rise to a ``resonant broadening effect'' \citep{1966PhFl....9.1773D} which produces a finite growth rate at $k_x = 5$. 

\begin{figure}
	\begin{centering}
		\includegraphics[width=\linewidth]{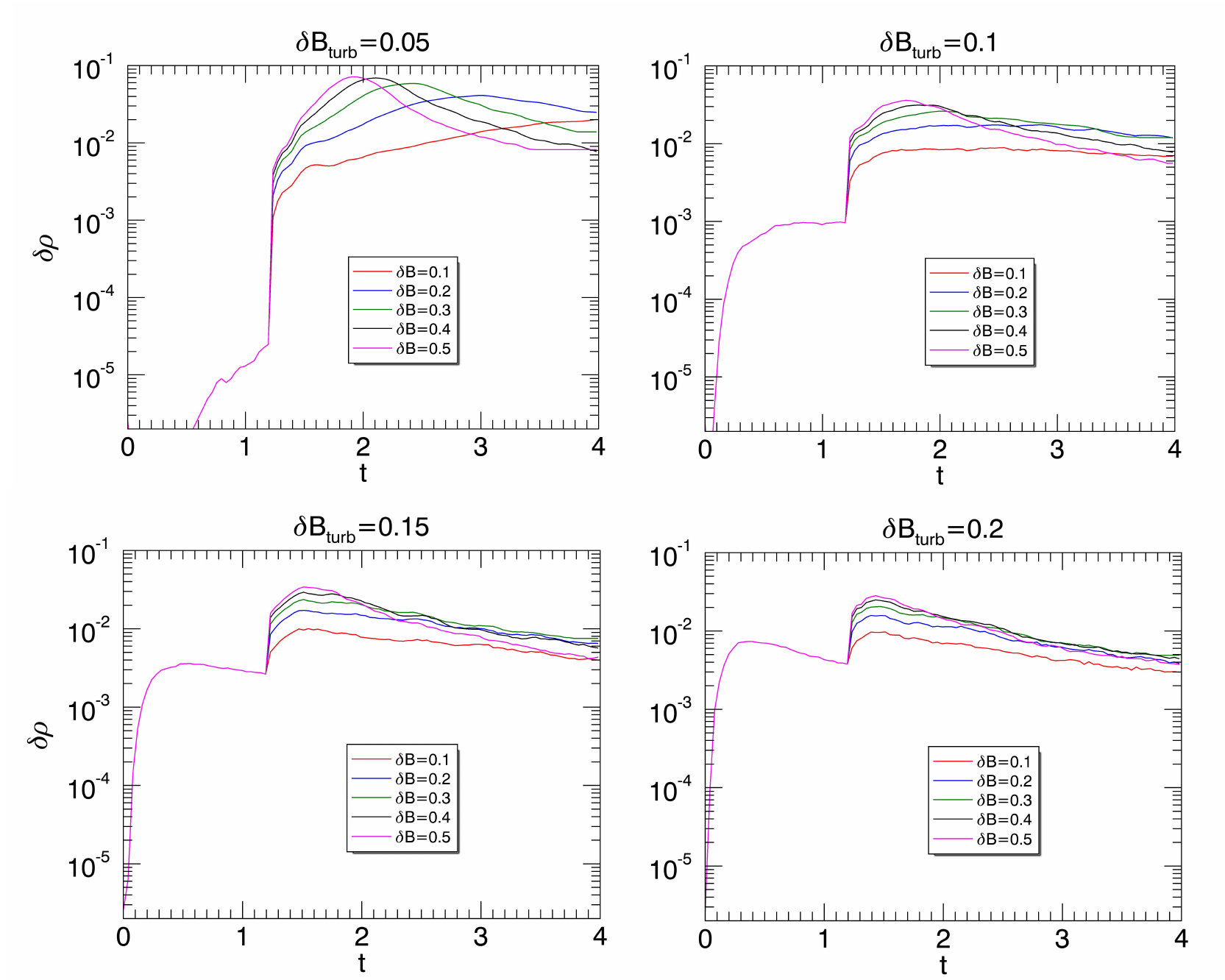}
		\caption{Evolution of PDI modes under different background turbulence levels ($\delta B_{turb}$) and different amplitudes of injected Alfv\'en waves ($\delta B$). A circularly polarized Alfv\'en wave is injected at $t= 1.2$.}
		\label{fg6}
	\end{centering}
\end{figure}

\begin{figure}
	\begin{centering}
		\includegraphics[width=\linewidth]{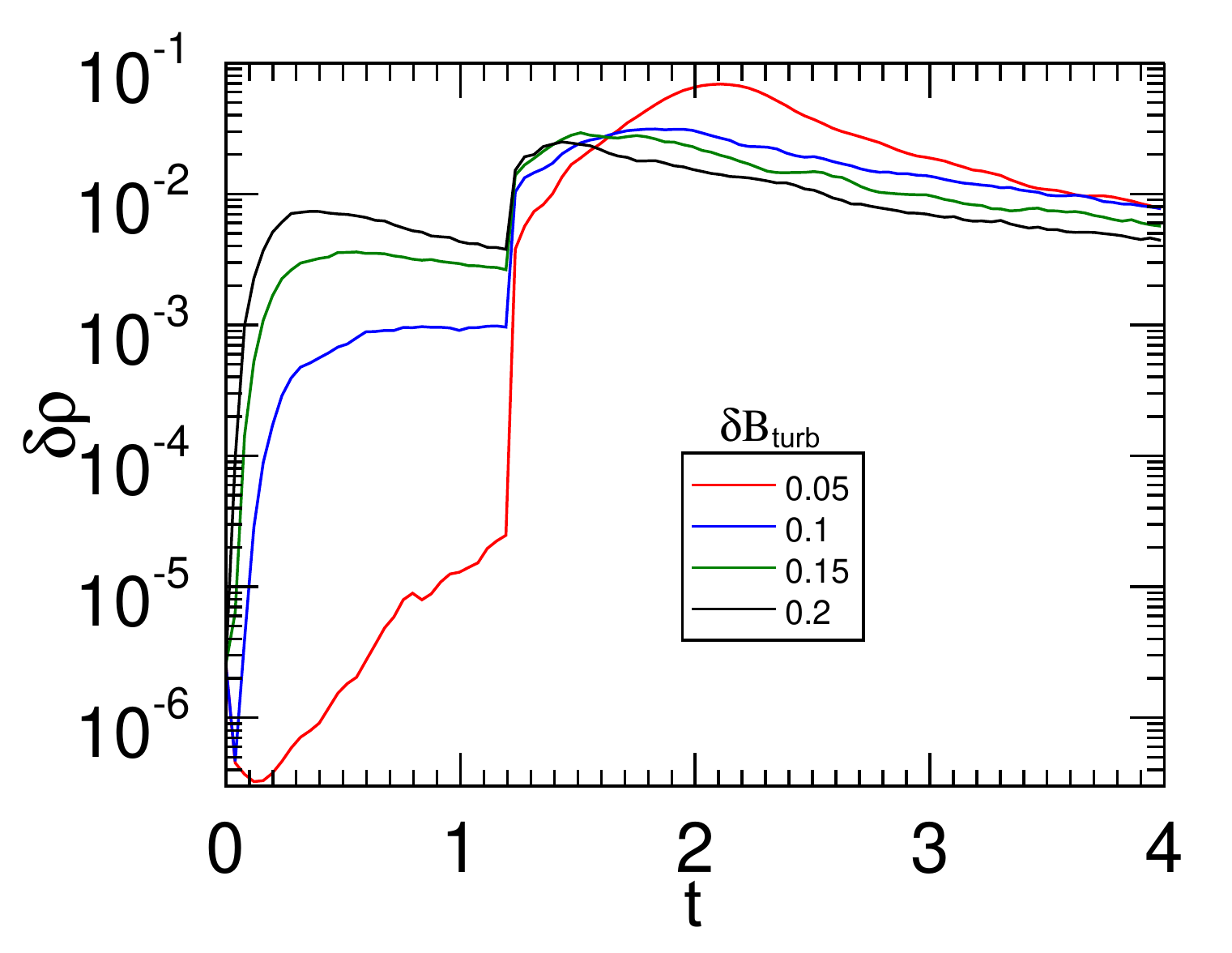}
		\caption{Evolution of PDI modes in different background turbulence levels ($\delta B_{turb}$). A circularly polarized Alfv\'en wave with $\delta B=0.4$ is injected at $t= 1.2$.}
		\label{fg7}
	\end{centering}
\end{figure}

Next, we show the evolution of PDI modes under different turbulence levels ($\delta B_{turb} = 0.05, 0.1,0.15, 0.2$)　and different amplitudes ($\delta B = 0.1$ to $0.5$) of injected Alfv\'en waves in Figure \ref{fg6}. While the existence and the growth rate of PDI are relatively easy to quantify in many runs, it becomes difficult when the turbulence level is high and the PDI stage is relatively short. Figure \ref{fg7} singles out the case with a fixed injected Alfv\'en wave amplitude ($\delta B=0.4$) while varying the background turbulence level. This figure gives a clearer sense on the effects of the background turbulence. Overall, we find several trends: 1. The background turbulence levels do influence the growth rate of PDI. The growth rate decreases with increasing turbulence background. It is difficult to quantify the reduction for all cases due to the relatively short growth stage of PDI with high turbulence levels. 2. The background turbulence levels do influence the saturation level of PDI. Higher turbulence background gives lower saturation level of density variations from PDI. 3. Higher turbulence level also shortens the duration of the PDI stage. 4. In all the runs we have made so far, PDI nonetheless still gets excited. This demonstrates the robustness of this instability.

\subsection{Generation of Slow Waves}

Density fluctuations arise during the growth of PDI of Alfv\'en waves. To understand the nature of these 
fluctuations better, we employ the spatial-temporal FFT analysis to determine whether the density variations 
follow dispersion relations. For example, using the 3D simulation shown in Figure \ref{fg2}(b) 
with $\delta B = 0.2$, we put 50 time frames of 3D data together between $t = 2$ and $2.4$
with a time resolution of $\delta t=0.008$ to form a big 4D matrix. 
The FFT analysis of the matrix gives power $P$ versus 
$\omega$ and $k$: $P(\omega,k_x,k_y,k_z)$. To get a clear view of the modes with the most power, 
we specify $k_y, k_z$ and get a contour figure in the $\omega - k_x$ plane. 
These modes are then compared with the linearized ideal MHD modes for the same choices of $k_y$ and $k_z$. 

%
%
%

\begin{figure}
	\begin{centering}
		\includegraphics[width=\linewidth]{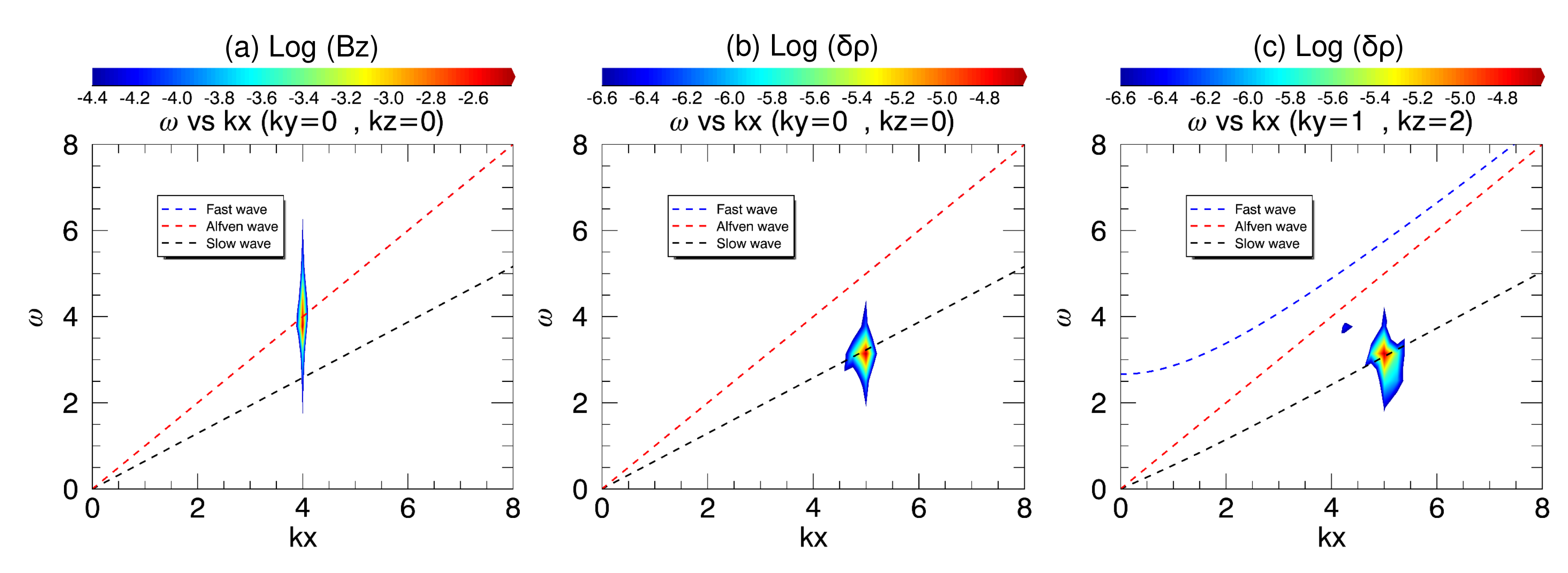}
		\caption{(a) Contour of mode ($k_x, k_y, k_z = 4, 0, 0$) from spatial-temporal FFT results of $B_z$ (the injected Alfv\'en wave), as well as the theoretical dispersion relations of fast (blue), Alfv\'en (red), and slow (black) waves. Note that fast wave and Alfv\'en wave are degenerate in this situation. (b) Contour of mode ($k_x, k_y, k_z = 5, 0, 0$) from spatial-temporal FFT results of $\delta\rho$, as well as the theoretical dispersion relations of fast (blue), Alfv\'en (red), and slow (black) waves. (c) Contour of mode ($k_x, k_y, k_z = 5, 1, 2$) from spatial-temporal FFT results of $\delta\rho$, as well as the theoretical dispersion relations of fast (blue), Alfv\'en (red), and slow (black) waves. Analyses uses the simulation data from $t=2$ and $2.4$, $\beta = 0.5$, and $\delta B = 0.2.$}
		\label{fg8}
	\end{centering}
\end{figure}

Figure \ref{fg8} (a) shows the contour figure of FFT results of $B_z$ in $\omega - k_x$ plane 
(with $k_y = 0, k_z = 0$), as well as the theoretical dispersion relation of three MHD waves. 
The  Alfv\'en wave dispersion curve crosses the high power region exactly, indicating that this mode  
($k_x = 4, k_y = k_z = 0$) is an Alfv\'en wave, which is actually the initial  circularly polarized pump 
Alfv\'en wave. This result confirms that our method is reliable. 
To analyze the nature of the compressible modes, we perform the spatial - temporal FFT of 
density fluctuations $\delta\rho$ during the same time domain. Figure \ref{fg8} (b) and (c) 
show the contours of two modes with high power, along with the theoretical dispersion relations of MHD waves. 
The mode in Figure \ref{fg8} (b) propagates parallel to the background magnetic field and is 
either a slow wave or a sound wave (they are degenerate in this situation). Figure \ref{fg8} (c) shows a mode with finite $k_y$ and $k_z$, and it 
matches well with the dispersion relation of a slow wave. 
The growth rate of the slow modes in Figure \ref{fg8} (b) and (c) is relatively large (see Figure \ref{fg4}). These results show that slow MHD waves are indeed generated via PDI.

\section{Summary and Conclusion}
\label{sec:sum} 

We have performed 3D ideal MHD simulations to study the parametric decay instability of 
a circularly polarized Alfv\'en wave in both the turbulent and non-turbulent environments, and 
compare the growth rate of PDI in different situations. We find that in the turbulence circumstance, 
the PDI of Alfv\'en waves still occurs, though their growth rate is reduced to be about 50\% 
of the growth rate predicted by the linear theory and confirmed by non-turbulence simulations. 
Because the growth time of the PDI modes typically last over $\sim 0.2-0.8~ \tau_A$, the injected
Alfv\'en wave could experience non-linear interactions with the background variations (such as 
Alfv\'en waves).  The reduced growth rate of PDI in turbulent background may be related to these
non-linear interactions, though further studies are needed to quantify this process. 
In addition, in certain cases, we find that the turbulence or background fluctuations can actually destabilize modes that are stable 
accordingly to the linear theory, an effect that seems to be similar to the resonance broadening process. 

To study the nature of the compressible fluctuations generated during the PDI of Alfv\'en waves, 
we use the 4D spatial-temporal FFT method to analyze the density and magnetic field fluctuations, 
and compare the dispersion relations from simulations with those from theoretical MHD waves. 
Our results show that slow MHD waves are generated during the PDI. We propose that 
this mechanism can explain some of recent observations of slow waves in the solar wind 
\citep[e.g.,][]{2012ApJ...753L..19H,2013ApJ...774...59Y,2015ApJ...813L..30H,2015ApJ...815..122S}. 

That slow waves can be produced by the PDI of Alfv\'en waves locally in the solar wind
may have significant implications to understanding the solar wind heating. 
In addition to the often-considered scenario of Alfv\'enic cascade transferring energy to small scales 
where dissipation occurs, Alfv\'en waves at large scales may continuously transfer energy to 
compressible modes via PDI on relatively large scales, then the dissipation of these compressible
modes via Landau damping can heat the solar wind plasmas directly. Recently, \cite{2016JPlPh..82b9012S} proposed an 'anti-phase-mixing' effect, which may suppress the damping and dissipation of compressible fluctuations in the turbulent background.
It will be quite interesting to carry out further studies of these processes. 

There are certain limitations of this work: We have used ideal MHD simulations that 
exclude Landau damping effect, which could affect the excitation of PDI in collisionless plasmas
such as in the solar wind \citep[e.g.,][]{2008PhRvL.100l5003A}. In addition, a continuously driven
turbulence situation is more realistic for studying the evolution of PDI in relation to the solar wind.
These effects are the subjects of future studies.

\acknowledgments
H. Li acknowledges useful discussions with B. Chandran. M. Shi thanks S. Li for his help in conducting MHD simulations. M. Shi and H. Li acknowledge the support by 
LANL/LDRD and DoE/OFES. Institutional computing resources were used for this study. 
This work is also supported by National Natural Science Foundation of China 
under 41274168, 40974104, 11375053 and ITER-CHINA program 2015GB120001, 2014GB107004. 
M. Shi is also supported by a scholarship from Graduate School of Peking University.

%




\bibliographystyle{aasjournal}
\bibliography{library}




\end{document}